

\input phyzzx

\hsize 6.5truein
\vsize 9.0truein
\hoffset 0.2truein
\voffset 0.06truein

\def\uk{\mu {\rm K}}
\def\singlespace{\baselineskip 12pt}

\def\medspace{\baselineskip 18pt}

\def\doublespace{\baselineskip 24pt}
\def\bk{\hfill\break}

\def\sqr#1#2{{\vcenter{\vbox{\hrule height.#2pt
       \hbox{\vrule width.#2pt height#1pt \kern#1pt
          \vrule width.#2pt}
       \hrule height.#2pt}}}}

\doublespace
\PhysRevtrue
\nopubblock

\REF\COBE{G. F. Smoot, {\it et al.} Astrophys. J.  Lett.
{\bf 396},
L1 (1992).}

\REF\GAIER{T. Gaier, {\it et al.} Astrophys. J. Lett. {\bf
398}, L1 (1992).}
\REF\SOUTHPOLE{P. Meinhold and P. Lubin, Astrophys. J.
Lett. {\bf 370},
11 (1991); P. Lubin, P. Meinhold, and A. Chingcuanco in
{\it
The Cosmic Microwave Background 25 Years Later}, edited by
N. Mandolesi
and N. Vittorio (Kluwer,Dodrecht, 1990).}
\REF\BELM{J. R. Bond, G. Efstathiou, P. M. Lubin, and P.
R. Meinhold,
Phys. Rev. Lett. {\bf 66}, 2179 (1991).}
\REF\GORSKI{For another analysis of the South Pole data
see K. M.
Gorski, R. Stompor, and R. Juszkiewicz, YITP preprint
92-36 (1992).}
\REF\BE{S. Dodelson and J. M. Jubas (in preparation).
Many of the methods we have used are taken from
J. R. Bond and G. Efstathiou, Mon. Not. Roy. Astron. Soc.
{\bf 226}, 655 (1987). For a clear review, see
G. Efstathiou in {\it Physics
of the Early Universe}, edited by J. A. Peacock, A. F.
Heavens, and
A. T. Davies (Edinburgh University Press, Edinburgh, 1990).}
\REF\PEEBYU{P. J. E. Peebles and J. T. Yu, Astrophys. J.
{\bf 162}, 815
(1970).}
\REF\WILSON{M. L. Wilson and J. Silk, Astrophys. J. {\bf
243}, 14 (1981).}
\REF\BSZ{J. R. Bond and A. S. Szalay, Astrophys. J. {\bf
274}, 443 (1983).}
\REF\READ{For a clear discussion of likelihoods, see
A. C. S. Readhead, {\it et al.}, Astrophys. J. {\bf 346},
566 (1989), section VIII.}
\REF\VITTORIO{N. Vittorio, P. R. Meinhold, P. F.
Muciaccia, P. M. Lubin,
and J. Silk, Astrophys. J. Lett. {\bf 372}, 1 (1991).}
\REF\ADAMS{The error here includes the effects of cosmic
variance.
For a clear discussion of how to normalize using COBE
data, see
G. Efstathiou, J. R. Bond, and S. D. M. White, Oxford
preprint (1992);
F. C. Adams, J. R. Bond, K. Freese, J. A. Frieman,
and A. V. Olinto, Phys. Rev. D (in press), section IV.}
\REF\STATS{Since $\chi^2$ for this data set is not small
[$5$ for $6$
degrees of freedom], we expect most other statistical
tests to give
similar answers. See Ref. \READ.}
\REF\CHITEST{Specifically we found that the $\chi^2$ for
the best fit
with $27$ degrees of freedom is $46$. Gaier, {\it et al.}
assumed a Gaussian
correlation function while we used the full correlation
function
induced by CDM.}
\REF\SILK{J. Silk, Astrophys. J. {\bf 151}, 459 (1968).}
\REF\SDAMP{At any given time, the Silk damping scale
is of order $(n_e \sigma_T H)^{-1/2}$. At decoupling then,
when
the free electron density $n_e \propto n_b^{1/2}$, the
Silk damping scale
[and hence the magnitude of this effect] remains constant
if
$\Omega_b h^4$ remains constant.}
\REF\BBN{S. Yang, M. S. Turner, G. Steigman, D. N.
Schramm, and K. A. Olive,
Astrophys. J. {\bf 281}, 493 (1984);
T. P. Walker, G. Steigman, D. N. Schramm, K. A. Olive,
and H. Kang, Astrophys. J. {\bf 376}, 51 (1991).}

\titlepage
\singlespace
\rightline{\hfill FERMILAB-Pub-92/366-A}
\rightline{\hfill December 1992}
\medspace

\title {Microwave Anisotropies in the Light of {\it COBE}}
\bk

\author {Scott Dodelson$^{1,}$\footnote{\sharp}{E-mail
address: Dodelson@fnal.fnal.gov} and Jay M.
Jubas$^{2,}$\footnote{\natural}{E-mail
address: jubas@pierre.mit.edu}}

\address {$^1$NASA/Fermilab Astrophysics Center\break
          Fermi National Accelerator Laboratory\break
          P.O. Box 500, Batavia, IL 60510 \break
                    \break
          $^2$Department of Physics \break
          Massachusetts Institute of Technology\break
          Cambridge, MA 02139}

\baselineskip 13pt
\vskip 0.4in
\centerline{ABSTRACT}
\vskip 0.25in

The recent COBE measurement of anisotropies in the cosmic
microwave
background and the recent South Pole experiment of Gaier
{\it et al.} offer an excellent opportunity to probe
cosmological
theories. We test a class of theories in which the
Universe today is flat and matter dominated, and
primordial perturbations
are adiabatic parameterized by an index $n$.
In this class of theories the predicted signal in
the South Pole experiment depends not only on $n$,
but also on the Hubble constant and the baryon
density. For $n=1$ a large region of this parameter space
is ruled
out, but there is still a window open which satisfies
constraints coming from
COBE, measurements of the age of the Universe, the South
Pole experiment,
and big bang nucleosynthesis. Using the central values of
the Hubble
constant and baryon density favored by nucleosynthesis and
age
measurements, we find that, even if the COBE normalization
drops by $1\sigma$, $n > 1.2$ is ruled out.
\endpage
\doublespace
\FIG\FLIKE{The likelihood function for the South Pole
experiment using
all four channels (dotted line) and only one channel
(solid lines). The
likelihood function has been normalized so that it is
equal to
$1$ at its peak.}
\FIG\FOMEGAH{Constraints from the South Pole experiment on
$h,\Omega_B$
assuming $n=1$. The region above the dashed line is ruled
out at the
$95\%$ confidence level if COBE normalization is used
$(Q=15\uk)$. The region allowed by
Big Bang Nucleosynthesis is bounded by solid lines.}
\FIG\FINDEX{Combined constraints on spectral index $n$ and
quadrupole $Q$ from COBE and Gaier {\it et al.} COBE allows the
region between the
solid lines [from a combination of their sky noise at
$10^\circ$ and
the full correlation function]. The Gaier experiment rules
out the
region above the dashed [short-dashed] line at the $95
(68) \%$
confidence level. Here we have set $h=.5$ and
$\Omega_b=.05$.}

The recent detection$^\COBE$
 by the {\it COBE} satellite of anisotropies
in the microwave background has important ramifications
for the ongoing
searches for anisotropies at smaller angular scales. In
particular, the {\it COBE} measurement can be used to normalize the
spectrum of  primordial
perturbations. This normalization, in any given theory,
gives an unambiguous
prediction for the magnitude of the anisotropy that should
be detected in
smaller scale experiments. Here we focus
on models in which the Universe is flat and matter
dominated and perturbations
are adiabatic and ask: Does
COBE's normalization of these theories imply that a signal
should have been seen in the smaller scale Gaier$^\GAIER$
experiment? Although our results have been
obtained by assuming cold dark matter (CDM), we expect
similar results for hot
dark matter or cold $+$ hot dark matter because the Gaier
experiment probes
scales so large that neutrino free streaming is
essentially irrelevant.

The South Pole
experiment$^{\GAIER,\SOUTHPOLE,\BELM,\GORSKI}$
 consists of a beam at fixed zenith angle
[$\theta_z = 27.75^\circ$] oscillating back and forth in
a given sky patch with period $1/\nu$. Thus the position
of the beam is determined by its
azimuthal angle: $\phi(t) = \phi_A \sin(2\pi\nu t)$; here
$\phi_A \sin\theta_z = 1.5^\circ$. When the beam gets halfway across
patch, the ``sign'' of the signal changes, so that the
expected signal is
$$ \delta T = 4\nu \int_{-\phi_A}^{\phi_A}
 {d\phi\over (d\phi/dt)}  S(\phi)
 T(\theta_z,\phi) \eqn\DTG$$
where $S$ is either
plus or minus one depending on the angle and $T$ is the
temperature.
It is customary to expand the temperature in multipole
moments so that
$T = T_0(1+ \sum_{l,m} a_{lm} Y_{lm})$, where $T_0$ is the
observed mean temperature
of the cosmic microwave background, $2.735^\circ K$, and
the $a_{lm}$ are Gaussian random variables. If many
measurements are
made, the mean value of the $a_{lm}$ should be zero, but
with a variance
given by $<a_{lm}^*a_{l'm'}> =
C_l\delta_{l,l'}\delta_{m,m'}$. After squaring
Eq. \DTG\ and inserting these relations, we see that a
cosmological theory which predicts a set of $C_l$'s
predicts a variance
in the Gaier experiment:
$$ < \left( {\delta T\over T_0} \right)_{th}^2 > =
	\sum_{l=2}^\infty {C_l\over 4\pi} (2l+1) W_l.
\eqn\DTTH$$
Here the filter function is
$$ W_l = \exp\left\{- (l+.5)^2 \theta_s^2 \right\}
	{16\pi\over 2l+1} \sum_{m=-l}^l H_0^2(m\phi_A)
Y_{lm}^2(\theta_z,0)
\eqn\FILTER $$
where $\theta_s = 0.425 \times 1.35^\circ$ represents the
width of the beam
and $H_0$ is the Struve function of order $0$. This filter
function
peaks at $l\sim 70$ and falls off significantly so that
the the contribution
from modes greater than $l\sim 250$ is negligible.
The Gaier experiment made
measurements over nine such patches in each of four
frequency channels.

To compare a given cosmological theory with the Gaier
experiment, therefore,
we must ask it for the $C_l$'s. For the adiabatic, matter
dominated
models under consideration, generating the $C_l$'s is
straightforward$^\BE$:
(i) perturb the Einstein and Boltzmann equations about the
standard zero order solutions [Robertson-Walker metric; homogeneous
and isotropic distributions of photons, neutrinos, ordinary
matter, and dark matter]$^{\PEEBYU,\WILSON,\BSZ}$; (ii) Fourier
transform these equations after which
the perturbations are functions of wavenumber $k$, time
$t$, and, in the
case of photons and neutrinos, the angle between the
wavenumber and momentum;
(iii) Expand the perturbations to the photons and
neutrinos in terms of
Legendre polynomials so that the angular dependence,
$\Delta(\mu)$, is replaced by the coefficients, $\Delta_l$; (iv) Evolve
these perturbed
quantities starting from initial conditions deep in the
radiation era:
$\delta\rho/\rho (k,t_{\rm init}) \propto k^{n/2}$ where
$n=1$ for
the Harrison-Zel'dovich spectrum predicted by inflation;
(v) Determine the $C_l$'s today by integrating $C_l
\propto \int d^3k \vert
\Delta_l (t_0)\vert^2$. The proportional signs in the
previous two sentences
show that these theories do not fix the normalization.
That is, there is no
prediction for a given $C_l$; however the ratio $C_l/C_2$
is unambiguously
determined. Therefore, the predicted signal in the Gaier
experiment,
$<\delta T_{th}^2>$,
depends on only one
parameter $C_2$, or equivalently the quadrupole $Q [=
\sqrt{5C_2/4\pi} T_0]$.

Let us take the quadrupole as a free parameter. Then in a
given patch
we can construct the probability density of a given
measurement [$\delta
 T_{obs}
\pm \sigma$]:
$$ P(\delta T_{obs} \vert Q) =  \left[ 2\pi (\sigma^2 +
<\delta T_{th}^2(Q)>)
\right]^{-1/2}\ \exp\left\{ {-\delta T_{obs}^2 \over
<\delta T_{th}^2(Q)>
+ \sigma^2} \right\}.   \eqn\LIKELI $$
Naively, if this probability density is significantly
lower at a value
of $Q$ than it is at its maximum, then we can confidently
rule out that
particular value of $Q$. The Gaier experiment has nine
patches, so
the nine probability densities must be multiplied together
to form the {\it likelihood function}$^\READ$. In fact, things are
a little
more complicated than this because the nine patches are
close to each other [in fact they overlap somewhat],
so that the expected signals in the nine patches are
correlated.
We have included
cross-correlations amongst the different patches;
this is a straightforward
extension of the above$^{\BELM,\VITTORIO}$.

Figure 1 shows the likelihood as a function of $Q$ for
several different
values of the Hubble constant [$H_0 = 100 h$ km sec$^{-1}$
Mpc$^{-1}$]
and baryon density [$\Omega_B$ is the ratio of the baryon
density
to the critical density]. While $Q=0$, corresponding to no
signal,
is the most likely value, clearly values of $Q$ up to
about $10 \uk$
are allowed. It is also clear that values of $Q$ greater
than about $20
\uk$ are ruled out. With this range in mind, we note that
the COBE-inferred  value of $Q$ is$^{\ADAMS}$
$15\pm 3 \uk$.

Is $Q=15 \uk$ ``ruled out'' by the Gaier experiment?
One way to answer this question$^{\STATS}$ is to perform a
Bayesian
analysis assuming a uniform prior$^\BELM$. All this means
here is we ask what
fraction of the area under the likelihood curve is taken
up by $Q>15$.
For $\Omega_B = .05; h=.05$, this fraction is only $4\% $,
so we
say that the theory is ruled out at the $96\% $ confidence
level.
However, this number becomes significantly less impressive
as the COBE normalization is lowered. $Q=12$, which is
allowed by  COBE
at the one sigma level is ``ruled out'' with only $91\%$
confidence.

Until now we have ignored the dotted line in Fig. 1;
the solid lines were
drawn using only the highest frequency channel from the
Gaier experiment. The lower three channels had larger
signals
[i.e. larger average values of $\vert \delta T\vert$]. The
dotted line in Fig. 1 shows what the likelihood function would
look like
if all four channels were included in the analysis. We see
that the most
likely value of the quadrupole is about $9 \uk$ and no
signal, or $Q=0$, is ruled out on the basis of the four
channel data!
The difference between analyzing all four channels of data
and analyzing only
the highest channel is immense: either we say that COBE
normalized
CDM is on the verge of being ruled out OR there has been a
detection
at roughly the level expected. The team analyzing the data
ran
extensive spectral tests and concluded that there is only
a $2\%$
probability
that the signal in the low channels is cosmic microwave
background.
[Other contributions, such as Bremsstrahlung and
synchrotron radiation,
 fall off as the frequency increases so the highest
channel
should be least contaminated by them.] We have run a
similar test$^{\CHITEST}$
and also find that the probability that the signal in the
four channels is pure cosmic background is very low. So we will follow
Gaier {\it et al.}
and consider only the highest channel of data in our
analysis.

The three solid lines in Fig. 1 make it clear that we
lied when we claimed that the signal expected from CDM
depends only
on the normalization. Clearly it depends on two other
parameters as well,
$h$ and $\Omega_b$. Figure 2 shows the allowed region of
parameter
space for $Q=15 \uk$, the central value of COBE.

There are two physical effects which lead to the shape of
this contour
plot.  The first effect relates to the
imperfect coupling between photons and baryons prior to
decoupling.
If the coupling were perfect,
the intrinsic photon fluctuations would be maximal, and
the the anisotropy in the photon
temperature would be quite large.
Since the coupling is not perfect, photons can diffuse out
of
perturbations, damping the temperature anisotropy [i.e.
the perturbations
undergo Silk damping$^{\SILK}$]. Therefore, the weaker
the interactions between photons and matter, the smaller
is the final
photon anisotropy. We know though that the interaction
rate
increases as the amount of matter increases.
Since the matter density scales as $\Omega_b h^2$, we
expect the
intrinsic anisotropy to {\it increase}$^{\SDAMP}$
as $h$ increases for fixed $\Omega_b$.
This effect shows up at the high $h$ end of Fig. 2, where
even relatively low values
of $\Omega_b$ are ruled out.
The second effect depends not on the matter, but rather
on the gravitational field through which the photons
travel before they reach us.
If $h$ is small, the Universe was {\it not} purely matter
dominated since the surface of last scattering.
The epoch at which the energy density in matter equals
that in radiation
comes closer to the epoch of last scattering as $h$
decreases, so that for at least
part of the photons' flight to us, the gravitational
potential was not
constant. This leads to an additional contribution to the
anisotropy
and hence a larger signal. This explains why for $h$ less
than $1/2$ or so, even small values of $\Omega_b$ are
ruled out.

Also shown in Fig. 2 is the allowed region in
$(h,\Omega_b)$ space
from primordial nucleosynthesis considerations$^{\BBN}$.
One might combine
the allowed BBN regime with the regime favored by
measurements of
the age of the Universe [e.g. restricting the age to be
greater than
$10$ billion years corresponds to $h<.65$] and direct
measurements of $h$ [which all observers would agree is greater than
$.4$].
While the Gaier experiment rules out
a large part of the $(h,\Omega_b)$ plane, it does {\it
not} rule out
this ``favored'' region of $h \sim .6$
and $\Omega_b \sim .03$.

What happens if the primordial spectrum differs from the
Harrison-Zel'dovich
spectrum predicted by inflation? Or, perhaps more to the
point, What
limits do microwave anisotropy experiments place on the
spectral index of
primordial perturbations? Figure 2 shows the values of the
 normalization $Q$ and spectral index $n$ allowed by COBE
and the
South Pole experiment. Large $n$ corresponds to more power
on small
scales and hence a larger predicted signal on angular
scales probed by Gaier, {\it et al.} Hence, the only way to
reconcile the
absence of a signal in the Gaier experiment with large $n$
is if the normalization $Q$ is small. For $n>1.2$ the upper
limit on $Q$
is smaller than the region favored by COBE.

To sum up our results: (i) The signal in medium scale
anisotropy
experiments depends not only on the assumed shape and
normalization
of the primordial spectrum but also on the Hubble constant
and the baryon
density; (ii) For CDM-like theories, the Gaier {\it et
al.} experiment, together with the normalization provided by
COBE,
rules out a large region of the $(h,\Omega_b)$ parameter
space; (iii)
There is still a window open which satisfies constraints
coming from
COBE, measurements of the age of the Universe, the Gaier
experiment,
and big bang nucleosynthesis; (iv) COBE and the Gaier
experiment
rule out values of the primordial spectral index $n >
1.2$.

\bk
\noindent
ACKNOWLEDGEMENTS

It is a pleasure to thank Ed Bertschinger and Albert
Stebbins for extensive
discussions about this work. We are also grateful to
Katherine Freese,
Todd Gaier, Josh Gundersen,
Steve Meyer, Michael Turner, and Martin White
for helpful comments.
The work of SD was supported in part by the DOE and NASA
grant
NAGW-2381 at Fermilab. JJ acknowledges the NSF under
Grant No. PHY-9296020 and the Sloan Foundation for partial
support.


\refout

\figout

\bye